\documentclass[12pt,epsf]{elsart}

\begin{document}

\begin{frontmatter}

\title{Predictions about the behaviour of diamond, silicon, SiC and 
some $A^{III}B^V$ semiconductor materials in hadron fields}

\author[univ]{I.Lazanu} and
\author[iftm]{S. Lazanu}

\address[univ]{University of Bucharest, Faculty of Physics,
P.O.Box  MG-11, Bucharest-Magurele, Romania,
electronic address: ilaz@scut.fizica.unibuc.ro}
\address[iftm]{National Institute for Materials Physics,
P.O.Box MG-7, Bucharest-Magurele, Romania, 
electronic address: lazanu@alpha1.infim.ro}

\begin{abstract}
The utilisation of crystalline semiconductor materials as detectors and 
devices operating in high radiation environments, at the future particle 
colliders, in space applications, in medicine and industry, makes 
necessary to obtain radiation harder materials. Diamond, SiC and 
different $A^{III}B^V$ compounds (GaAs, GaP, InP, InAs, InSb) are possible 
competitors for silicon to different electronic devices for the 
up-mentioned applications. The main goal of this paper is to give 
theoretical predictions about the behaviour of these semiconductors in 
hadron fields (pions, protons). The effects of the interaction between 
the incident particle and the semiconductor are characterised in the 
present paper both from the point of view of the projectile, the 
relevant quantity being the energy loss by nuclear interactions, and of 
the target, using the concentration of primary radiation induced defects 
on unit particle fluence. Some predictions about the damage induced by 
hadrons in these materials in possible applications in particle physics 
and space experiments are done.


\medskip
\end{abstract}

\end{frontmatter}

\section{Introduction}
The crystalline materials for semiconductor devices used in high 
fluences of particles are strongly affected by the effects of radiation. 
After the interaction between the incoming particle and the target, 
mainly two classes of degradation effects are observed: surface and bulk 
material damage, the last due to the displacement of atoms from their 
sites in the lattice. After lepton irradiation, the effects are 
dominantly at the surface, while heavy particles (hadrons and ions) 
produce both types of damages.

Up to now, in spite of the experimental and theoretical efforts, the 
understanding of the behaviour of semiconductor materials in radiation 
fields, the identification of the induced defects and their 
characterisation, as well as the explanation of the degradation 
mechanisms are still open problems.

The utilisation of semiconductor materials as detectors and devices 
operating in high radiation environments, at the future particle 
colliders, in space applications, in medicine and industry, makes 
necessary to obtain radiation harder materials. 

Diamond, SiC and different $A^{III}B^V$ compounds (GaAs, GaP, InP, InAs, 
InSb) are in principle, possible competitors for silicon in the 
realisation of different electronic devices. 

All analysed materials have a zinc-blend crystalline structure, with the 
exception of SiC, that presents the property of polytypism \cite{Verma}. The polytypism refers to one-dimensional polymorphism, 
i.e. the existence of different stackings of the basic structural 
elements along one direction. More than 200 polytypes have been reported 
in literature \cite{Lambrecht}, but only few of them have 
practical importance. These include the cubic form $3C (\beta)$, and the $4H$
and $6H$ 
hexagonal forms. For the cubic polytype, the symmetry group is $T_d^2$, while 
for the hexagonal ones this is $C_{6v}^4$. Silicon is at the base of electronic 
industry, diamond and the $A^{III}B^V$ compounds present attractive electrical 
and/or luminescence properties, of interest for different applications, 
while the utilisation of SiC as a radiation detector, both in high 
energy physics and in the field of X-ray astronomy is now under 
extensive investigation more conferences in the field having sections 
dedicated to SiC.

The diamond has the reputation of being a radiation hard material and it 
is considered as a good competitor to silicon, but non all its properties 
as a radiation hard material have been proved experimentally. 

The main goal of this paper is to give some theoretical predictions 
about the behaviour of different semiconductors in hadron fields (pions, 
protons), these materials representing potential candidates for 
detectors and electronic devices working in hostile environments.

The treatment of the interaction between the incident particle and the 
solid can be performed from the point of view of the projectile or of 
the target. In the first case, the relevant quantity is the energy loss 
(or equivalently the stopping power) and in the second situation the 
effects of the interactions are described by different physical 
quantities characterising material degradation. There is no a physical 
quantity dedicated to the global characterisation of the effects of 
radiation in the semiconductor material. A possible choice is the 
concentration of primary radiation induced defects on the unit particle 
fluence (CPD), introduced by \cite{Lazanu 1998a}. It permits the 
correlation of damages produced in different materials at the same 
kinetic energy of the incident hadron. For the comparison of the effects 
of different particles in the same semiconductor material, the non 
ionising energy loss (NIEL) is useful. 

As a measure of the degradation to radiation, in the present paper the 
energy lost by the incident particle in the nuclear interaction and the 
concentration of primary defects induced in semiconductor bulk are 
calculated. If the energy loss of the incident projectile is, in 
principle, a measurable physical quantity, the concentration of primary 
defects is not directly observable and measurable and can be put in 
evidence only indirectly, e.g. from the variation of macroscopical 
parameters of the material or/and of electronic devices. It is to be 
noted that there exists also  a kinetics of the defects induced by 
irradiation giving rise to the annealing process. A general treatment of 
the evolution processes is not possible, so, only some particular models 
exist in literature, see for example \cite{Lazanu 2000} and the 
references cited therein in the case of silicon. In these circumstances, 
in the present paper only the primary process of defect generation is 
modelled .This way, the energy range of incident hadrons for which the 
concentration of primary defects do not affect irreversibly the device 
properties could be established (theoretically predicted).

\section{Model of the degradation}
\subsection{Energy loss}
At the passage of the incident charged particle in the semiconductor 
material same of its energy is deposited into the target. The charged 
particles interact with both atomic and electronic systems in a solid. 
The total rate of energy loss, could, in general, be divided 
artificially into two components, the nuclear and the electronic part.

The energy lost due to interactions with the electrons of the target 
gives rise to material ionisation, while the energy lost in interactions 
with nuclei is at the origin of defect creation.

A comprehensive theoretical treatment of electronic stopping, which 
covers all energies of interest, cannot be formulated simply because of 
different approximations concerning both the scattering and contribution 
of different electrons in the solid. For fast particles with velocities 
higher then the orbital velocities of electrons, the Bethe-Bloch formula 
is to be used \cite{Caso}. At lower velocities, inner electrons 
have velocities greater than particle velocity, and therefore do not 
contribute to the energy loss. This regime has been modelled for the 
general case by Lindhard and Scharff\cite{Lindhard 1961} and particular
cases have
been treated, e.g. in reference \cite{Morvan}. If the particle has a 
positive charge, and a velocity close to the orbital velocity of its 
outer electrons, it has a high probability of capturing an electron from 
one of the atoms of the medium through which it passes. This process 
contributes to the total inelastic energy loss since the moving ion has 
to expend energy in the removal of the electrons which it captures.

The nuclear stopping depends on the detailed nature of the atomic 
scattering, and this in turn depends intimately on the form of the 
interaction potential. At low energies, a realistic potential based on 
the Thomas-Fermi approximation has been used
in the literature \cite{Lindhard 1961}
and at higher energies, where scattering results from the
interaction of unscreened nuclei, a Rutherford collision model is to be 
used.

\subsection{Bulk defect production}
The mechanism considered in the study of the interaction between the 
incoming particle and the solid, by which bulk defects are produced, is 
the following: the particle, heavier than the electron, with electrical 
charge or not, interacts with the electrons and with the nuclei of the 
crystalline lattice. The nuclear interaction produces bulk defects. As a 
result of the interaction, depending on the energy and on the nature of 
the incident particle, one or more light particles are produced, and 
usually one or more heavy recoil nuclei. These nuclei have charge and 
mass numbers lower or at least equal to those of the medium. After the 
interaction process, the recoil nucleus or nuclei, if they have 
sufficient energy, are displaced from the lattice positions into 
interstitials. Then, the primary knock-on nucleus, if its energy is 
large enough, can produce the displacement of a new nucleus, and the 
process could continue as a cascade, until the energy of the nucleus 
becomes lower than the threshold for atomic displacements. 

The concentration of the primary radiation induced defects on unit 
fluence has been calculated starting from the following equation:

\begin{equation}
CPD\left( E\right) =\frac 1{2E_d}\int \sum_{\mbox{i}} \frac{d\sigma _i}{d\Omega }%
L\left( E_{Ri}\right) d\Omega  \nonumber
\end{equation}

where $E$ is the kinetic energy of the incident particle, $E_d$ the threshold
energy for displacements in the lattice, $E_{Ri}$ the recoil energy of the
residual nucleus, $L\left( E_{Ri}\right) $ the Lindhard factor describing
the partition between ionisation and displacements and $d\sigma _i/d\Omega $
the differential cross section for the process responsible in defect
production. In the concrete calculations, all nuclear processes, and all
mechanisms inside each process are included in the summation over index $i$.
Because of the regular nature of the crystalline lattice, the 
displacement energy is anisotropic. 

In the concrete evaluation of defect production, the nuclear 
interactions must be modelled, see for example references
\cite{Lazanu 1998a,Lazanu 1997,Lazanu 1999a,Lazanu 1998b}. The primary
interaction between the hadron and the nucleus of the lattice presents 
characteristics reflecting the peculiarities of the hadron, especially 
at relatively low energies. If the inelastic process is initiated by 
nucleons, the identity of the incoming projectile is lost, and the 
creation of secondary particles is associated with energy exchanges 
which are of the order of MeV or larger. For pion nucleus processes, the 
absorption, the process by which the pion disappears as a real particle, 
is also possible.

The energy dependence of cross sections, for proton and pion interaction 
with the nucleus, presents very different behaviours: the proton-nucleus 
cross sections decrease with the increase of the projectile energy, then 
have a minimum at relatively low energies, followed by a smooth increase,
while the pion nucleus cross sections present for all processes a large
maximum, at about 160 MeV,
reflecting the resonant structure of interaction (the $\Delta _{33}$ resonance
production), followed by other resonances, at higher energies, but with
much less importance. Due to the multitude of open channels in these
processes, some simplifying hypothesis have been done \cite{Lazanu 1998b}.

The process of partitioning the energy of the recoil nuclei (produced 
due the interaction of the incident particle with the nucleus, placed in 
its lattice site) by new interaction processes, between electrons 
(ionisation) and atomic motion (displacements) is considered in the 
frame of the Lindhard theory \cite{Lindhard 1963}.

The factor characterising recoil energy partition between ionisation 
and displacements has been calculated analytically, solving the general 
equations of the Lindhard theory in some physical approximations. 
Details about the hypothesis used could be found in reference
\cite{Lazanu 1999b}. All curves start, at low energies, from the same 
curve; they have at low energies identical values of the energy spent 
into displacements, independent on the charge and mass number of the 
recoil. At higher energies, the curves start to detach from this main 
branch. This happens at lower energies if their charge and mass numbers 
are smaller. The maximum energy transferred into displacements 
corresponds to recoils of maximum possible charge and mass numbers. The 
curves present then a smooth increase with the energy. For the energy 
range considered here, the asymptotic limit of the displacement energy 
is not reached.

For binary compounds, the Lindhard curves have been calculated 
separately for each component of the material, and the average weight 
Bragg additivity has been used. In this case, a threshold for atomic 
displacements must be considered for each atomic species and for each 
direction in the crystal. In the concrete calculations, a weighted 
value, independent on the crystalline direction has been used.

\section{Results, discussions and some possible applications}
The nuclear stopping power presents an energy dependence with a 
pronounced maximum. It is greater for heavier incident particles: 
protons compared to pions. In a given medium, the position of the 
maximum is the same for all particles with the same charge. In figure 1, 
the nuclear energy loss in diamond, silicon, silicon carbide, GaP, GaAs, 
InP, InAs and InSb is represented for protons and pions respectively, as 
a function of their kinetic energy. In the same medium, the position of 
the maximum is the same for pions and protons.

The behaviour of these materials in proton and pion fields is 
characterised by the CPD. In Figure 2, the dependence of the CPD as a 
function of the protons kinetic energy and medium mass number is 
presented for diamond, silicon, SiC GaAs and InP - see reference 
\cite{Lazanu LB 2000} and references cited therein. The values for 
diamond degradation are from reference \cite{Lazanu 1999a}, 
the corresponding ones for silicon are averaged values from 
references \cite{Ginneken} and \cite{Summers},
SiC - from reference
\cite{Lazanu Ae 2000} and those for GaAs and InP are from
reference \cite{Summers}. Low kinetic energy protons produce 
higher degradation in all materials. The discontinuity in the surface is 
related to differences in the behaviour of the CPD for monoatomic 
materials (or binary ones with close elements), and binary ones with 
remote elements in the periodic table.

For pion induced degradation, the energy dependence of CPD (as well as 
of the NIEL) presents two maxima, the relative importance of which 
depends on the target mass number: one in the region of the $\Delta _{33}$ 
resonance, more pronounced for light elements and compounds containing 
light elements, and another one around 1 GeV kinetic energy, more 
pronounced for heavy elements. At higher energies, an weak energy 
dependence is observed, and a general $A_{average}^{3/2}$ dependence of the NIEL can be 
approximated \cite{Lazanu 1998b,Lazanu 1999b}.
In Figure 3, the CPD for all analysed materials
(diamond, Si, SiC, GaP, GaAs, InP, InAs, InSb) is represented as a 
function of the pion kinetic energy and of material average mass number. 
The differences in the behaviour of these materials are clearly 
suggested by the discontinuity in the mesh surfaces.

In the energy range considered in the paper, it could be observed that 
the CPD produced by pions and protons, and characterising the bulk 
degradation, are very different and reflect the peculiarities of the 
interactions of the two particles with the semiconductors. For pions, 
there are two maxima, one in the region of the $\Delta _{33}$ resonance, 
corresponding to about 140 - 160 MeV kinetic energy, and the other at 
higher energies, around 1 GeV. The relative importance of these maxima 
depends on the mass number of the material. In comparison with this 
behaviour, the CPD produced by protons decreases abruptly with the 
increase of energy at low energies, followed by a smooth and slow 
increase at higher energies.

In relation to their behaviour in pion fields, these materials could be 
separated into two classes, the first with monoatomic materials or 
materials with relatively close mass numbers (diamond, silicon, GaAs and 
InSb), and the second comprising binary materials with remote mass 
numbers of the elements (SiC, GaP, InP, InAs) with similar behaviours 
inside each group. The diamond is the hardest material from all 
considered here. A slow variation of the primary defect concentration 
has been found for pion irradiation of diamond, silicon, SiC, GaP and 
GaAs, in the whole energy range of interest, with less than 2 
displacements/cm/unit of fluence. In contrast to these materials, there 
are others, characterised by a low CPD in the energy range up to 200 MeV, (which represents this way the upper limit of the energy range where their utilisation in pion field is recommended), followed by a pronounced increase of displacement concentration with energy to more than 8 displacements/cm/unit fluence for InSb. 

It is to be mentioned that, in the present model hypothesis, for SiC, 
negligible differences have been found between 
different polytypes in what regards the effects of pion and proton 
degradation \cite{Lazanu Ae 2000}, conclusions in accord with the
experimental results
\cite{Wendler}. The behaviour of SiC in radiation 
fields is between the corresponding one of diamond and silicon.

As it is well known, the analysed semiconductors are possible materials 
for detectors and electronic devices which have to work long time in 
particle physics experiments, space applications, etc., in intense 
fields of hadrons, and in experimental configurations which impose high 
reliability of devices, and must present a controlled degradation of 
their parameters. As possible applications we will analyse two 
hypothetical cases: the utilisation of diamond, silicon, SiC or GaAs as 
detectors at the Large Hadron Collider (LHC) at CERN, and the long time 
exposure of the electronic devices in the field produced by cosmic 
rays.

For the LHC, the standard physics programme is based on the study of  
proton - proton interactions, at  about 7 TeV beam energy, on an 
integrated luminosity of $5x10^5$ pb$^{-1}$ which corresponds to 9 year of 
operation, for an annual operation time of 1.9x10$^7$ s. The irradiation 
background  is continuous. The charged hadrons are produced in the 
primary interactions, while the neutrons are albedo particles. 
The charged pions are the dominant particles, followed by protons, 
antiprotons and kaons.  As an illustration of the above calculations 
for the degradation of different semiconductors in proton and pion 
fields, the results of the simulation of Gorfine and Taylor \cite{Gorfine} have 
been chosen, for pion and proton fluxes in the inner detector assembly 
region, parallel to the beam axis. Both protons and pions are 
transported down to thermal energies in the detector, by nuclear 
interactions. The particle flux energy spectra have been simulated for a 
first layer of Si detectors (situated at 11.5 cm), and with complete 
moderator. The obtained spectra have been found  to be slowly dependent 
on the material of the inner detector and so, in the present paper the 
same hadron spectra for diamond, silicon carbide, silicon and gallium 
arsenide have been utilised. 

The convolution of the pion and proton spectra with the energy 
dependence of the CPD has been done in the energy range 50 MeV - 10 GeV, 
and 10 MeV - 10 GeV for pions and protons respectively. Below 50 MeV, a 
realistic estimation of materials degradation to pions is very difficult 
due to the lack of experimental data on pion - nucleus interaction and 
also to the increase of the weight of Coulomb interaction.

The results of these calculations are summarised in Fig. 4, for the 
diamond,  silicon carbide, silicon and GaAs options, both for pion and 
proton degradation. In the analysed case, diamond and SiC are the 
hardest materials in both pion and proton fields; the diamond is harder 
to pions than to protons. The behaviour of silicon is similar in both 
particle fields. The GaAs option is not recommended because of an order 
of magnitude higher degradation in comparation with all other considered 
materials.

Another possible utilisation of semiconductor devices in radiation 
fields is related to space applications. In the primary cosmic radiation, 
the most abundant particles are the protons \cite{Caso}. Other 
charged particles (for example $\pi^{+/-}$, $e^{+/-}$, $\mu^{+/-}$,
$\nu_{mu}$, etc.) are produced in the
interaction of the primary cosmic rays in air. The damage induced in 
diamond, Si, SiC, GaAs and InP by the primary cosmic field has been 
estimated for protons in the energy range 10 MeV - 10 GeV, and the 
results are presented in Figure 5. The devices have been supposed to be 
exposed directly to the cosmic field. In this case too, diamond has been 
found to be the hardest material. GaAs and InP suffer a degradation of a 
factor of about 50 times higher in comparation with diamond and this 
behaviour can affect irreversibly the properties of these materials for 
long time operation.

The degradation produced by the particle field at LHC (pions and 
protons) and by the free protons from the primary cosmic rays 
respectively are of the same order of magnitude for each of the 
materials investigated.

\section{Summary}

A systematic theoretical study has been performed, investigating the 
interaction of charged hadrons with semiconductor materials and the 
mechanisms of defect creation by irradiation. 

The nuclear stopping power has been found to be greater for heavier 
incident particles (protons compared to pions), and for lighter media. 
The position of its maximum is the same for protons and pions in the 
same medium.

The mechanisms of the primary interaction of the hadron with the 
nucleus (nuclei) of the semiconductor lattice have been explicitly 
modelled and the Lindhard theory of the partition between ionisation 
and displacements has been  applied.

For protons, the low kinetic energy particles produce higher 
degradation in all materials.

For pions, the energy dependence of CPD presents two maxima, the 
relative importance of which depends on the target mass number: one in 
the region of the $\Delta _{33}$ resonance, more pronounced for light
elements and
compounds containing light elements, and another one around 1 GeV 
kinetic energy, more pronounced for heavy elements. At higher energies, 
an weak energy dependence is observed. A slow variation of the primary 
defect concentration has been found for pion irradiation of diamond, 
silicon, GaP and GaAs, in the whole energy range of interest, with less 
than 2 displacements/cm/unit of fluence. In contrast to this situation, 
for the other semiconductor materials analysed, a low CPD is estimated 
in the energy range up to 200 MeV (which represent the energy range up 
to their utilisation in pion field is recommended), followed by a 
pronounced increase of displacement concentration to more than to more 
than 8 displacements/cm/unit of fluence at high energies.

The behaviour of this semiconductor materials has been analysed 
comparatively both in relation to particle physics experiments (inner 
part of the detection system at LHC) and to space applications (the 
devices being considered to be exposed directly to the cosmic ray 
field.

\newpage

\begin{center}
\bf{Figure captions}
\end{center}
\bigskip
\medskip

Figure 1:
The nuclear energy loss in diamond, silicon, silicon carbide, GaP, GaAs, 
InP, InAs and InSb
as a function of the kinetic energy of the incident particle: protons 
(up) and pions (down) respectively.

\medskip

Figure 2a:  
The concentration of primary defects on unit  fluence (CPD) in  diamond,
silicon,  SiC,
GaAs and InP induced by protons, as a function of the kinetic energy and
average mass number of the semiconductor material. The mesh surfaces are
drawn only to guide the eye.

Figure 2b:
The dependence of CPD as a function of proton kinetic energy, for the same
semiconductors.

\medskip

Figure 3a:
The energy  and  material   dependence of  the CPD  on  unit  pion  
fluence  for diamond, Si,
SiC, GaP, GaAs, InP, InAs and InSb. The mesh surfaces are drawn to guide 
the eyes.

Figure 3b:
The CPD as a function of  the kinetic
energy of incident pions, for the same semiconductors.

\medskip

Figure 4:
Estimated CPD on unit fluence induced in diamond, silicon, SiC and GaAs, 
by the simulated
flux energy spectra of pions and protons in the inner detector at LHC.

\medskip

Figure 5:
Estimated  CPD  on  unit  fluence  induced  by  the primary cosmic ray 
flux energy  spectra in
diamond, Si, SiC, GaAs and InP (only the effects produced by protons are 
considered), in the hypothesis that these semiconductor materials are 
exposed directly in the radiation field.

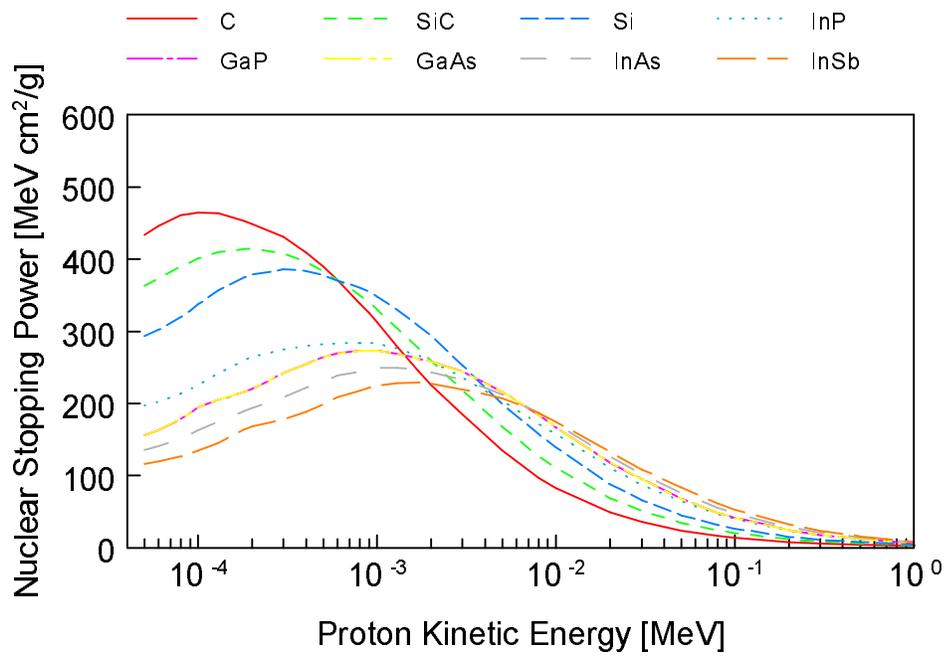
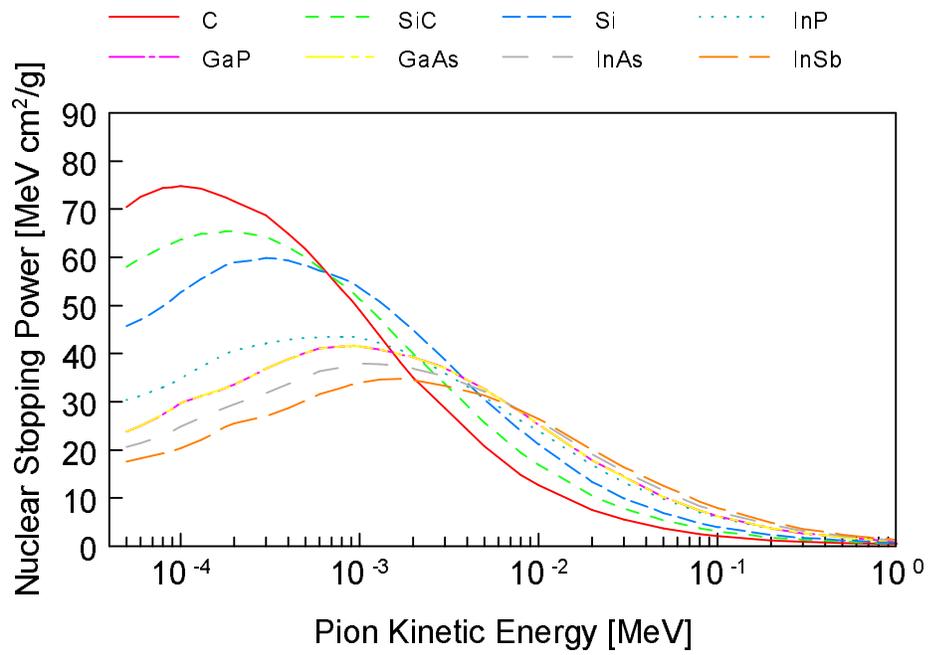
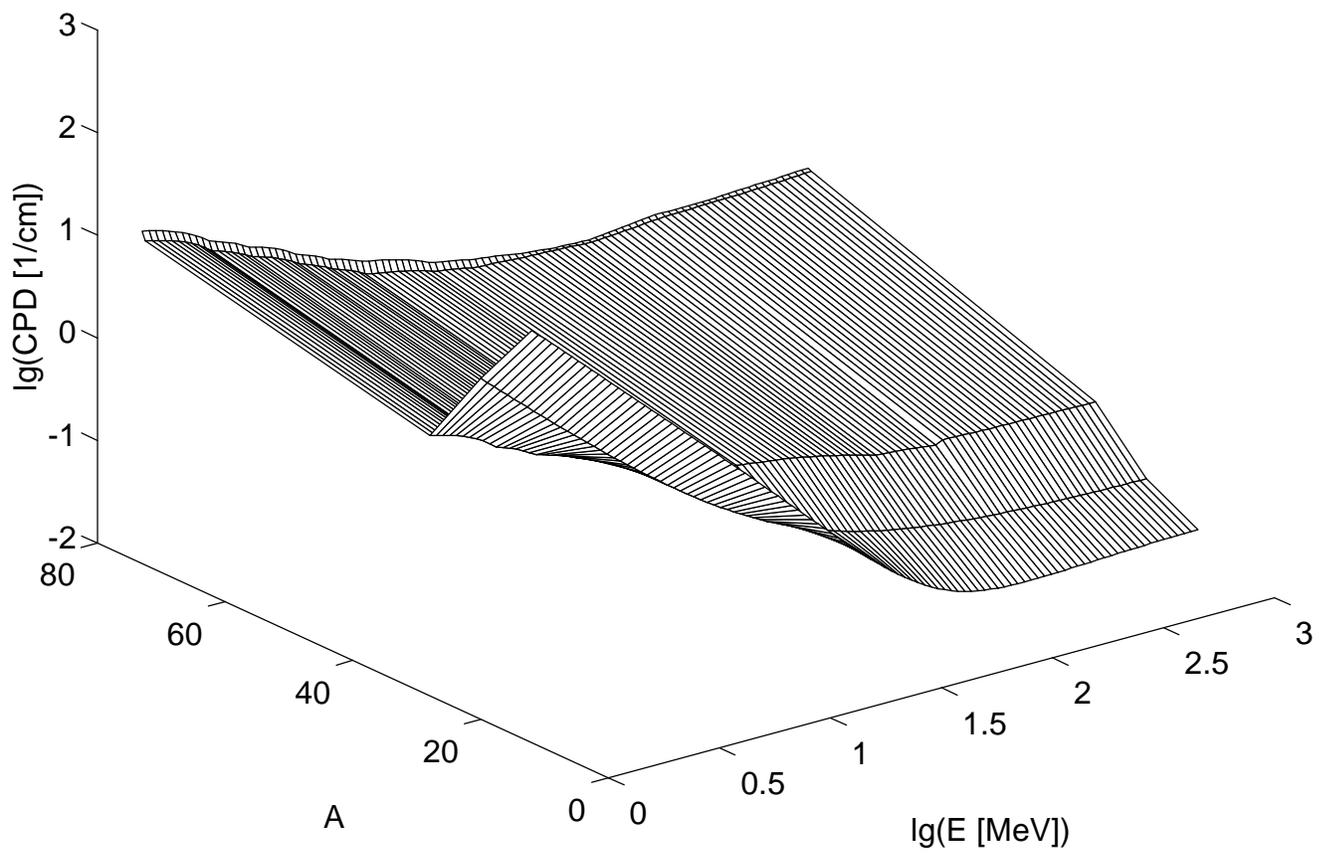
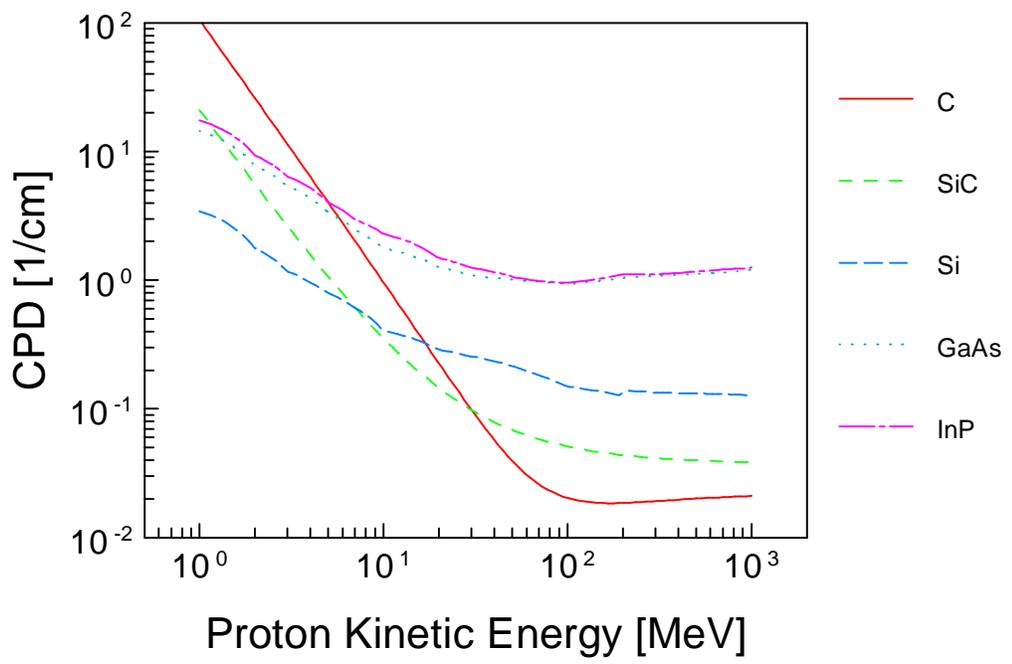
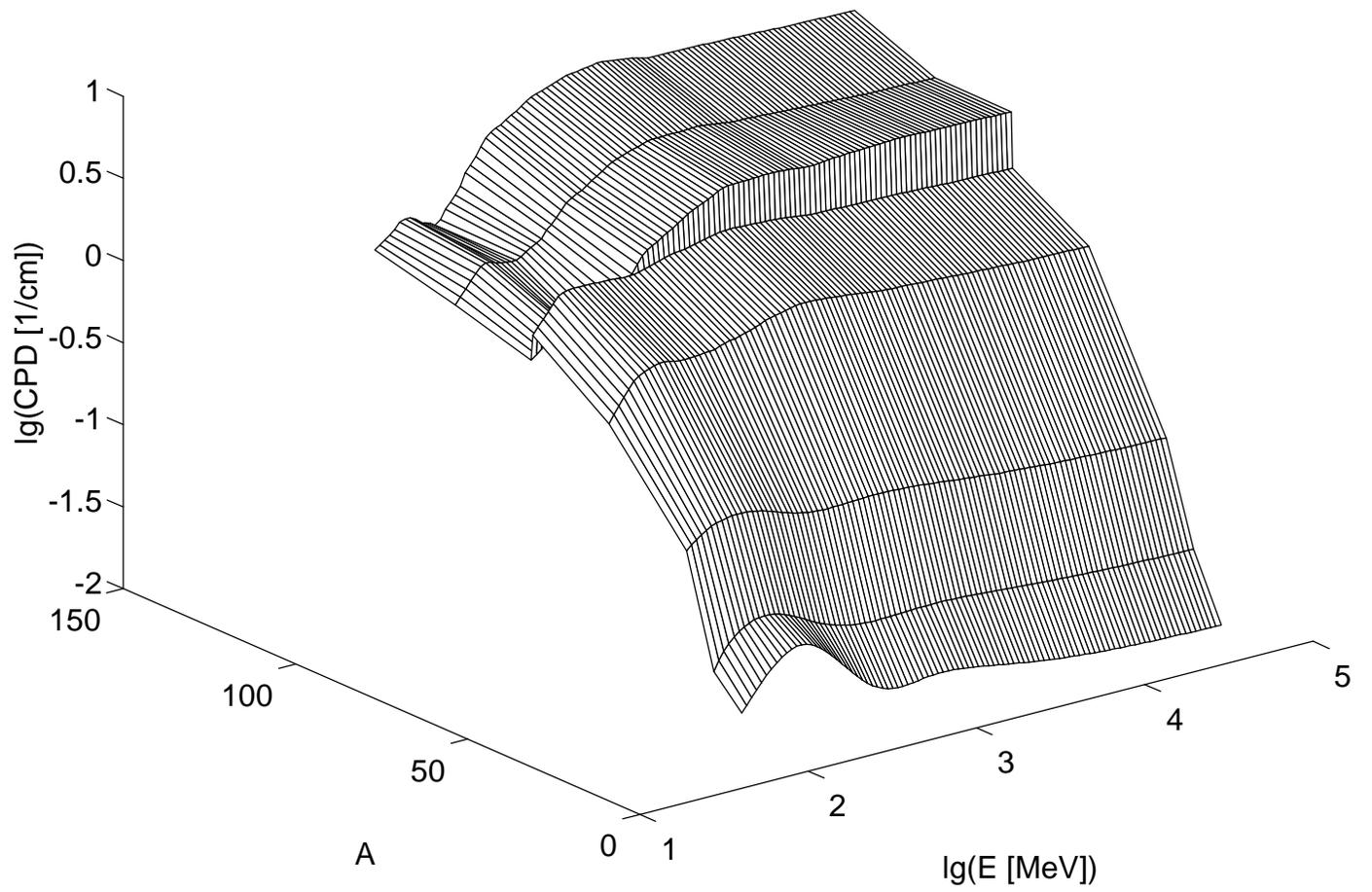
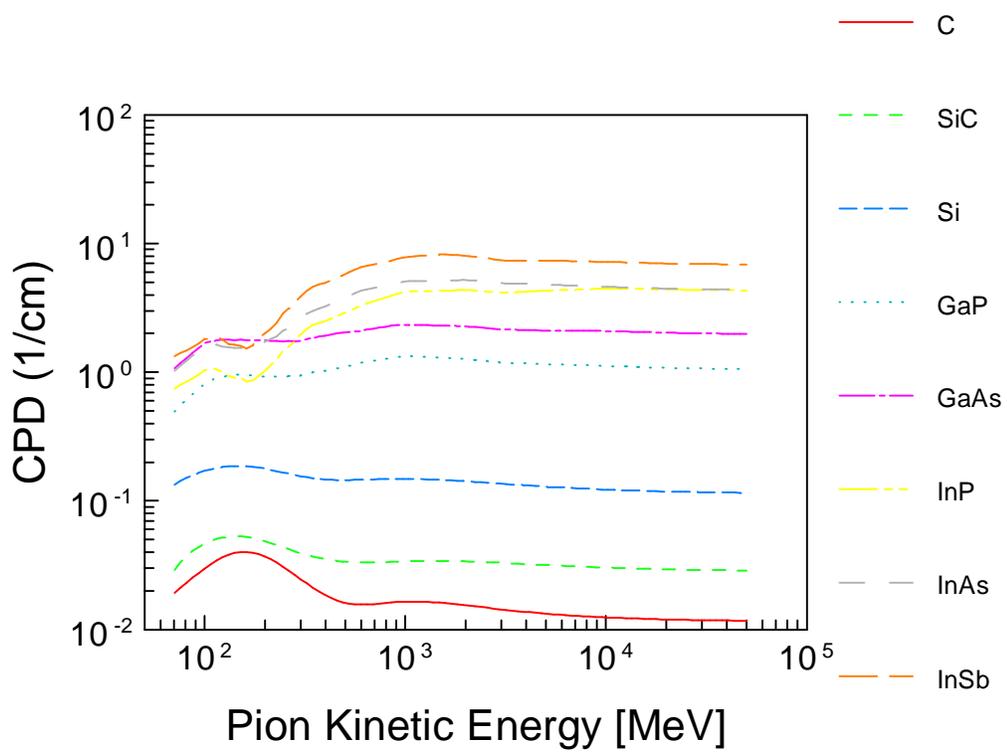
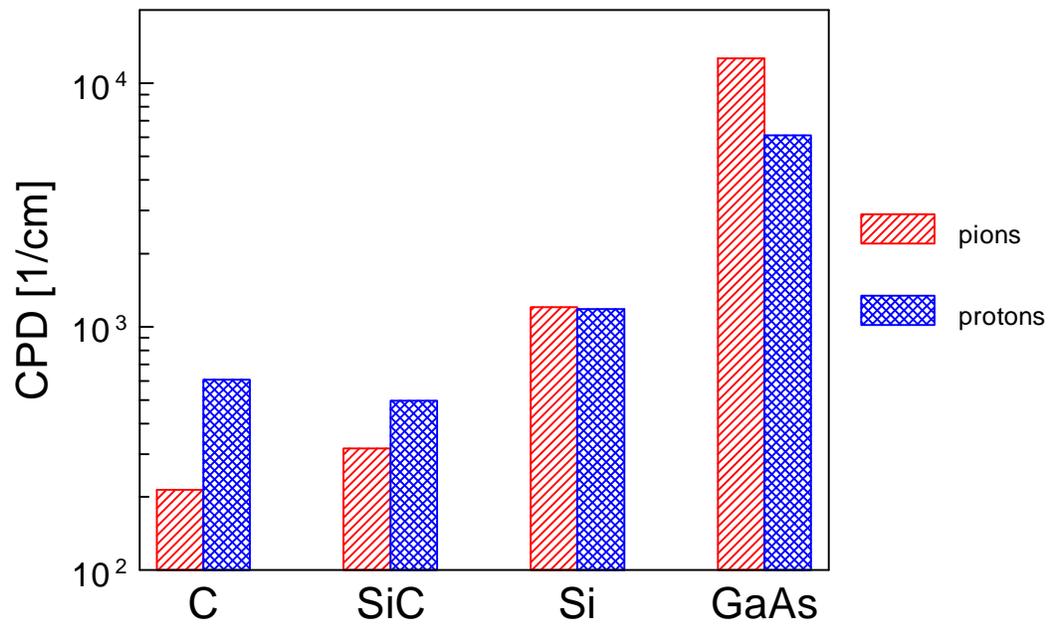
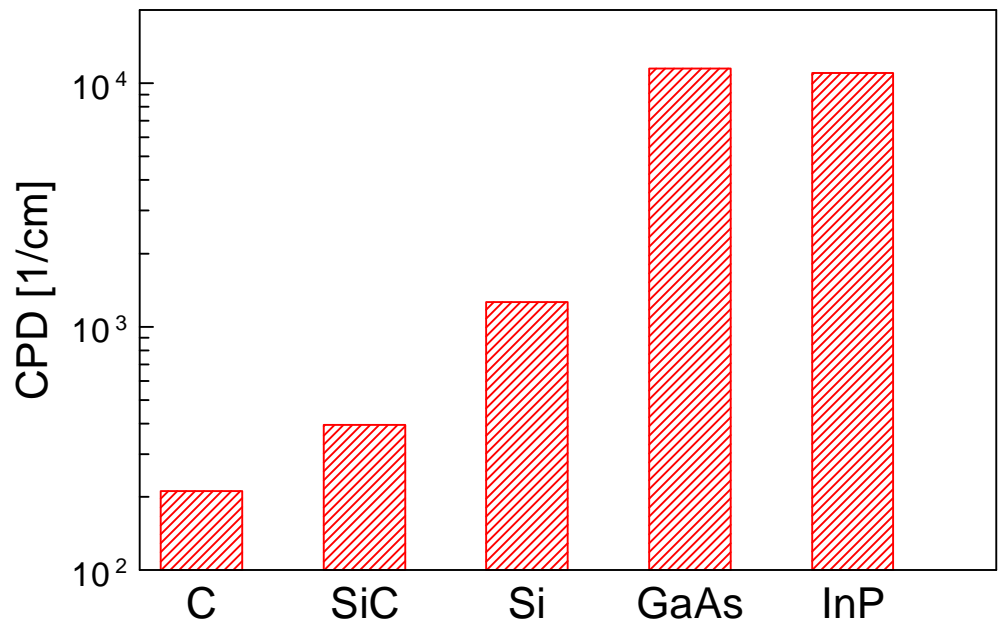
\end{document}